\documentclass[fleqn,usenatbib]{mnras}
\usepackage{newtxtext,newtxmath}
\usepackage[T1]{fontenc}

\DeclareRobustCommand{\VAN}[3]{#2}
\let\VANthebibliography\thebibliography
\def\thebibliography{\DeclareRobustCommand{\VAN}[3]{##3}\VANthebibliography}

\usepackage{orcidlink}
\usepackage{graphicx}
\usepackage{amsmath}
\usepackage{multirow}

\title[EP250108a/SN\,2025kg: A GRB--SN from Binary]{EP250108a/SN\,2025kg: A Magnetar-powered Gamma-Ray Burst Supernova Originating from a Close Helium-star Binary via Isolated Binary Evolution} 

\author[Zhu et al.]{
Jin-Ping Zhu\orcidlink{0000-0002-9195-4904},$^{1,2,3,4}$\thanks{E-mail: \url{jpzhu.astro@gmail.com}}
Jian-He Zheng\orcidlink{0000-0001-5751-633X},$^{5,6}$\thanks{E-mail: \url{mg21260020@smail.nju.edu.cn}}
Bing Zhang\orcidlink{0000-0002-9725-2524}$^{1,2,7,8}$\thanks{E-mail: \url{bzhang1@hku.hk}}
\\
$^{1}$The Hong Kong Institute for Astronomy and Astrophysics, The University of Hong Kong, Pokfulam Road, Hong Kong, People's Republic of China\\
$^{2}$Department of Physics, The University of Hong Kong, Pok Fu Lam, Hong Kong, People's Republic of China\\
$^{3}$School of Physics and Astronomy, Monash University, Clayton Victoria 3800, Australia\\
$^{4}$OzGrav: The ARC Centre of Excellence for Gravitational Wave Discovery, Clayton Victoria 3800, Australia\\
$^{5}$School of Astronomy and Space Science, Nanjing University, Nanjing 210023, People's Republic of China\\
$^{6}$Departments of Astronomy and Theoretical Astrophysics Center, UC Berkeley, Berkeley, CA 94720, USA\\
$^{7}$Nevada Center for Astrophysics and Department of Physics and Astronomy, University of Nevada, Las Vegas, NV 89154, USA \\
$^8$Department of Physics and Astronomy, University of Nevada, Las Vegas, NV 89154, USA
}

\date{Accepted xxx. Received xxx; in original form xxx}
\pubyear{2025}

\begin{document}
\label{firstpage}
\pagerange{\pageref{firstpage}--\pageref{lastpage}}
\maketitle

\begin{abstract}

SN\,2025kg, linked to EP250108a, is among the brightest broad-lined Type Ic supernova (SN Ic-BL) known, showing unique helium absorptions, a late-time broad H$\alpha$, and an early bump. In this {\em{Letter}}, we propose a jet-cocoon origin to explain EP250108a as off-axis cooling emission from a mildly relativistic inner cocoon viewed at $\sim45^\circ$ and the early bump of SN\,2025kg as the outer cocoon cooling emission, both constraining an energy of $\sim(1-2)\times10^{52}{\rm{erg}}$ and a progenitor radius of $\sim5\,R_\odot$. To explain SN\,2025kg's exceptionally luminous peak, potential energy injection into the $\sim2.5\,M_\odot$ ejecta from a magnetar with initial period $\sim1.7\,{\rm{ms}}$ and magnetic field $\sim2\times10^{15}{\rm{G}}$ may be required, implying a rapidly rotating $\sim4\,M_\odot$ progenitor. Thus, the progenitor may be a low-mass helium star with an extended helium envelope, supported by helium absorption lines and an inferred weak pre-SN wind. Hydrogen-rich material may reside in the inner ejecta layers, as suggested by the late-time broad H$\alpha$, possibly originating from main-sequence companion material evaporated by the magnetar wind. Since the observed near-solar metallicity challenges the popular quasi-chemically homogeneous evolution channel, the rapidly rotating helium-star progenitor of EP250108a/SN\,2025kg might attain angular momentum by being tidally spun up by a main-sequence companion in a close binary formed through isolated binary evolution.

\end{abstract}


\section{Introduction} \label{sec:Intro}

The progenitors of long-duration gamma-ray bursts (LGRBs) and their associated broad-lined Type Ic supernovae (SNe Ic-BL) are still not well constrained. LGRBs and SNe Ic-BL are generally believed to be powered by post-collapse central engines, such as energy injection from an accreting, rapidly spinning black hole \citep{Woosley1993,Popham1999} or a spinning-down millisecond magnetar \citep{Usov1992,Dai1998,Wheeler2000,Zhang2001}, implying progenitors with exceptionally high angular momentum. These rapidly rotating progenitors are commonly thought to be compact carbon-oxygen Wolf-Rayet (WR) stars that have lost hydrogen and/or helium envelopes. Popular formation scenarios include rapidly rotating, low-metallicity stars that undergo quasi-chemically homogeneous evolution (CHE) driven by rotational mixing \citep[e.g.,][]{Yoon2006,Woosley2006,Cantiello2007}, eventually producing WR stars. Alternatively, these rapidly rotating WR stars may originate from close binary systems through isolated binary evolution, where the stars can be tidally spun up by companions \citep[e.g.,][]{Izzard2004,Fuller2022,Hu2023}. 

The GRB jet launched from the central engine interacts with the stellar envelope, forming a jet head composed of a forward shock into the stellar material and a reverse shock into the jet \citep{RamirezRuiz2002,Matzner2003,Zhang2003,Bromberg2011}. The shocked stellar material and shocked jet within the jet head flow sideways, producing an outer cocoon and an inner cocoon, respectively, which wrap around the jet and influence its propagation. Cocoons can produce rapidly evolving, luminous thermal cooling emission and nonthermal afterglow emission following breakout, which typically appears before the SN peak \citep{Nakar2017,Piro2018,DeColle2018,DeColle2022}, as plausibly evidenced by the early-time bumps observed in XRF\,060218/SN\,2006aj \citep{Campana2006,Soderberg2006,Pian2006}, GRB\,171205a/SN\,2017iuk \citep{Izzo2019}, and EP240414a/SN\,2024gsa \citep{Sun2024,Bright2025,Srivastav2025,vanDalen2025}. Such pre-SN emission produced by the cocoon and SN emission may carry information about the progenitor’s radius and mass, providing valuable constraints on its physical properties. For instance, \cite{Zheng2025} interpreted the prompt emission of EP240414a as cooling emission from the inner cocoon and suggested that the early bump included a component of outer cocoon cooling emission, both of which constrained the progenitor’s radius to be $\sim3\,R_\odot$ \citep[cf.][]{Hamidani2025b,Hamidani2025a}.

Most recently, \cite{Li2025} reported a fast X-ray transient at $z=0.176$ {(a luminosity distance of $D_{\rm L}=881\,{\rm Mpc}$)}, EP250108a, characterized by a peak energy of $<1.8\,{\rm keV}$ and an isotropic-equivalent energy of $<10^{49}{\rm erg}$, with X-ray properties similar to those of XRF\,060218. SN\,2025kg, associated with EP250108a, is among the brightest SNe Ic-BL detected to date \citep{Li2025,Srinivasaragavan2025}. Detailed observations also revealed several unique spectral features for SN\,2025kg, such as He \textsc{i} absorption lines and a late-time broad H$\alpha$ line \citep{Rastinejad2025}, which have not been identified in other SNe Ic-BL. Furthermore, a unique early bump was detected in SN\,2025kg before the peak, most likely caused by the cooling emission \citep{Li2025,EylesFerris2025,Srinivasaragavan2025}. The progenitor and the explanations for these unique observational signatures of this source are still under debate. In this {\em Letter}, we propose a plausible scenario in which EP250108a/SN\,2025kg is interpreted as a magnetar-powered SN Ic-BL associated with a moderately off-axis ($\sim45^\circ$) jet–cocoon system originating from an expanded helium star in a close binary with a main-sequence (MS) companion. 

\section{Magnetar-powered SN Emission and Cocoon Cooling Emissions} \label{sec:AnalyticalEstimate}

\subsection{SN\,2025kg is Possibly a Magnetar-powered SN Ic-BL}

\cite{Li2025} revealed that SN\,2025kg peaks at $\sim14.5\,{\rm days}$ with a peak luminosity of $\sim2\times10^{43}\,{\rm erg}\,{\rm s}^{-1}$. Lightcurve modeling by \cite{Li2025,Srinivasaragavan2025} infers an ejecta mass of $M_{\rm ej}\sim2\,M_\odot$ and a kinetic energy of $E_{\rm ej,kin}\sim10^{52}{\rm erg}$, respectively. Following the \cite{Arnett1982} law, the corresponding SN peak time can be approximated by the ejecta diffusion timescale: 
\begin{equation}
    t_{\rm ej,diff}\approx\left(\frac{3\kappa M_{\rm ej}}{4\pi v_{\rm ej,f}c}\right)^{1/2}\approx14\,E_{\rm ej,kin,52}^{-1/4}\left(\frac{M_{\rm ej}}{2\,M_\odot}\right)^{3/4}{\rm days},
\end{equation}
consistent with the observed peak time, where the final velocity is $v_{\rm ej,f}\approx\sqrt{2E_{\rm ej,kin}/M_{\rm ej}}$ and the opacity is set to $\kappa\approx0.1\,{\rm cm}^2{\rm g}^{-1}$ \citep[e.g.,][]{Inserra2013}. Hereafter, the conventional notation $Q_x=Q/10^x$ is adopted in cgs units. The peak luminosity is approximately equal to the instantaneous injected luminosity at peak time. If powered by the radioactive decay of $^{56}$Ni and $^{56}$Co, with the radioactivity luminosity given by $L_{\rm Ni}(t)=(3.24e^{-t/\tau_{\rm Ni}}+0.73e^{-t/\tau_{\rm Co}})\times10^{10}M_{\rm Ni}\,{\rm erg}\,{\rm s}^{-1}$ \citep{Nadyozhin1994}, the SN peak luminosity can be estimated as
\begin{equation}
\begin{split}
   & L_{\rm ej,peak} \approx L_{\rm Ni}(t_{\rm ej,diff})\\
    &\approx 2\times10^{43}\left(\frac{M_{\rm Ni}}{0.8\,M_\odot}\right)\left(\frac{e^{-t_{\rm ej,diff}/\tau_{\rm Ni}}+0.2e^{-t_{\rm ej,diff}/\tau_{\rm Co}}}{0.4}\right){\rm erg}\,{\rm s}^{-1},
\end{split}
\end{equation}
where $M_{\rm Ni}$ is the $^{56}$Ni mass, $\tau_{\rm Ni}=8.8\,{\rm days}$, and $\tau_{\rm Co}=111.3\,{\rm days}$ respectively. Therefore, the observed peak properties suggest $M_{\rm Ni}\sim0.8\,M_\odot$ and $M_{\rm ej}\sim2\,M_\odot$. The inferred $^{56}$Ni mass fraction of SN\,2025kg, $f_{\rm Ni}=M_{\rm Ni}/M_{\rm ej}\sim0.4$, is much higher than the typical observed values of $f_{\rm Ni}\lesssim0.1$ from lightcurve fittings of a sample of SNe Ic-BL \citep[e.g.,][]{Lv2018,Taddia2019}, as well as the recent simulated range of $f_{\rm Ni}\sim0.025-0.3$ predicted by the black hole collapsar scenario \citep{Fujibayashi2024,Shibata2025}. These differences imply that radioactive decay alone may be insufficient to explain the exceptionally bright peak of SN\,2025kg.

We then consider the possibility that the peak of SN\,2025kg is powered by a newborn magnetar. Assuming that the energy injection is dominated by magnetic dipole spin-down radiation from the magnetar, with an initial spin period of $P_{\rm i}$ and a pole strength of $B_{\rm p}$, the spin-down luminosity is $L_{\rm sd}(t)=L_{\rm sd,i}(1+t/t_{\rm sd})^{-2}$, where $L_{\rm sd,i}\simeq E_{\rm rot}/t_{\rm sd}=B_{\rm p}^2R_{\rm mag}^6\Omega_{\rm i}^4/6c^3\simeq10^{47}B_{\rm p,14}^2P_{\rm i,-3}^{-4}\,{\rm erg}\,{\rm s}^{-1}$ is the initial luminosity, $t_{\rm sd}=3c^3I_{\rm mag}/B_{\rm p}^2R_{\rm mag}^6\Omega_{\rm i}^2\simeq2\times10^5\,B_{\rm p,14}^{-2}P_{\rm i,-3}^2\,{\rm s}$ is the spin-down timescale, $E_{\rm rot}=I_{\rm mag}\Omega_{\rm i}^2/2\simeq2\times10^{52}P_{\rm i,-3}^{-2}\,{\rm erg}$ is the rotational energy, $R_{\rm mag}\simeq10^6{\rm cm}$ is the magnetar's radius, and $I_{\rm mag}\simeq10^{45}{\rm g}\,{\rm cm}^2$ is the moment of inertia, respectively. LGRB and SN Ic-BL magnetars typically have observed magnetic fields of $\gtrsim4\times10^{14}{\rm G}$ \citep[e.g.,][Zhu 2025, in prep]{Lv2018} and hence have a very short spin-down timescale with $t_{\rm sd}\ll t_{\rm ej,diff}$, indicating that most of the rotational energy is deposited into kinetic energy, with only $\lesssim1-2\%$ converted into thermal emission. Based on the inferred kinetic energy of SN\,2025kg, the magnetar's initial spin period is $P_{\rm i}\sim1.4\,E_{\rm ej,kin,52}^{1/2}\,{\rm ms}$. Then, assuming $B_{\rm p}\sim1.2\times10^{15}{\rm G}$ and hence $t_{\rm sd}\sim2.7\times10^3B_{\rm p,15.1}^{-2}P_{\rm i,-2.85}^{2}\,{\rm s}$, the SN peak luminosity is
\begin{equation}
\begin{split}
    L_{\rm ej,peak}&\approx L_{\rm sd}(t_{\rm ej,diff})\approx L_{\rm sd,i}t_{\rm sd}^2/t_{\rm ej,diff}^2\\
    &= 2\times10^{43} B_{\rm p,15.1}^{-2}P_{\rm i,-2.85}^{-1}\left(\frac{M_{\rm ej}}{2\,M_\odot}\right)^{-3/2}{\rm erg}\,{\rm s}^{-1},
\end{split}
\end{equation}
in good agreement with the observed peak luminosity of SN\,2025kg. Thus, we propose that the peak of SN\,2025kg is possibly powered by a spin-down magnetar with $P_{\rm i}\sim1.4\,{\rm ms}$ and $B_{\rm p}\sim1.2\times10^{15}{\rm G}$.

\subsection{Jet Dynamics and Cooling Emissions from Cocoons}

{Apart from the long-term magnetic dipole decay powering the SN explosion, some short-term processes in the newborn magnetar, such as magnetic dissipation due to differential rotation \citep{Kluzniak1998} or accretion \citep{ZhangD2010}, may drive relativistic GRB jets.} Within the stellar envelope, the jet head typically propagates at a non-relativistic velocity, with $\beta_{\rm h} \simeq \tilde{L}^{1/2}$. The critical parameter $\tilde{L}$, which represents the dimensionless ratio of the jet's energy density to the rest-mass energy density of the stellar envelope materials, is given by $\tilde{L}\approx (L_{\rm j}/2\rho_\star t_{\rm b}^2\theta_{\rm j}^4c^5)^{2/5}$, where $L_{\rm j}$ is the total two-sided jet luminosity, $\rho_\star\approx3M_{\rm ej}/4\pi r_\star^3$ is the characteristic density of the stellar envelope, $r_\star$ is the progenitor's radius, $t_{\rm b}\sim r_\star/\beta_{\rm h}c$ is the jet breakout timescale, and $\theta_{\rm j}$ is the initial jet's half-opening angle. While simulations typically assume progenitor stars with masses around $\sim10\,M_\odot$ \citep[e.g.,][]{Mizuta2013,Nakar2017}, EP250108/SN\,2025kg appears to originate from a lower-mass progenitor, estimated to be $\sim4\,M_\odot$ if the remnant is an NS, potentially leading to a larger stellar radius and hence a wider jet half-opening angle $\theta_{\rm j}$. We conservatively adopt a larger value $\theta_{\rm j}\sim15^\circ$ instead of the canonical $\sim10^\circ$. Then, the jet breakout timescale becomes $t_{\rm b}\approx(3\theta_{\rm j}^4M_{\rm ej}r_\star^2/2\pi L_{\rm j})^{1/3}\approx4\,L_{\rm j,51}^{-1/3}\theta_{\rm j,-0.6}^{4/3}r_{\star,11}^{2/3}(M_{\rm ej}/2\,M_\odot)^{1/3}{\rm s}$, and the cocoon energy is $E_{\rm c}\approx L_{\rm j}t_{\rm b}\approx4\times10^{51}L_{\rm j,51}^{2/3}\theta_{\rm j,-0.6}^{4/3}r_{\star,11}^{2/3}(M_{\rm ej}/2\,M_\odot)^{1/3}{\rm erg}$.  Since the inner and outer cocoons typically share comparable energies after jet breakout, the energy deposited into both components of EP250108/SN\,2025kg is likely several times $10^{51}-10^{52}{\rm erg}$. {The inner and outer cocoons, after the breakout, can move relativistically and transrelativistically, respectively, leading to X-ray and ultraviolet-optical-infrared cooling emissions (see Figure \ref{fig:Illustration} in Appendix \ref{app_sec:Illustration} for the schematic illustration).}

\subsubsection{Pre-SN Bump from Outer Cocoon Cooling Emission} \label{sec:OuterCocoonModel}

Here, we explore whether the early-time bump observed before the main emission of SN\,2025kg can be attributed to the outer cocoon cooling emission basically following the analytic framework in \cite{Nakar2017,Piro2018}. The outer cocoon inside the progenitor is roughly a cylinder with a two-sided volume of $\sim2\pi \theta_{\rm j}^2r_\star^3$ and hence the mass of $M_{\rm oc}\sim 3\theta_{\rm j}^2M_{\rm ej}/2\approx0.2\,\theta_{\rm j,-0.6}^2(M_{\rm ej}/2\,M_\odot)\,M_\odot$. Once the jet breaks out, the outer cocoon undergoes rapid expansion, during which most of its internal energy is transformed into kinetic energy. We parameterize the kinetic energy distribution as a self-similar function of the velocity $v_{\rm oc}$, with $dE_{\rm oc,kin}/dv_{\rm oc}\propto v_{\rm oc}^{-s}$, corresponding to a mass profile of
\begin{equation}
\label{equ:MassDistribution}
    m_{\rm oc}(>v_{\rm oc})\approx(3\theta_{\rm j}^2M_{\rm ej}/2)\cdot(v_{\rm oc}/v_{\rm oc,min})^{-s-1},
\end{equation}
where $v_{\rm oc,min}\approx\sqrt{2E_{\rm oc}/M_{\rm oc}}=0.2\,E_{\rm oc,52}^{1/2}\theta_{\rm j,-0.6}^{-1}(M_{\rm ej}/2\,M_\odot)^{-1/2}c$ is the minimum velocity of the outer cocoon.

{Following \cite{Nakar2010}, the emission from a given velocity shell becomes visible to an observer when its optical depth drops to $\tau_{\rm oc}(>v_{\rm oc,diff})\sim\kappa m_{\rm oc}(>v_{\rm oc,diff})/4\pi v_{\rm oc,diff}^2t^2 \sim c/v_{\rm oc,diff}$, where $c$ is the speed of light. Solving for $v_{\rm oc,diff}$ yields the velocity of the diffusion shell, and substituting this $v_{\rm oc,diff}$ into Equation (\ref{equ:MassDistribution}), we obtain the mass of the material contributing to the diffusion emission as $m_{\rm oc,diff}\approx(3\theta_{\rm j}^2M_{\rm ej}/2)\cdot(t/t_{\rm oc,diff})^{2(s+1)/(s+2)}$,} where
\begin{equation}
    t_{\rm oc,diff}=\left(\frac{3\theta_{\rm j}^2\kappa M_{\rm ej}}{8\pi cv_{\rm oc,min}}\right)^{1/2}\approx 1.4\,E_{\rm oc,52}^{-1/4}\theta_{\rm j,-0.6}^{3/2}\left(\frac{M_{\rm ej}}{2\,M_\odot}\right)^{3/4}{\rm days}
\end{equation}
is the characteristic diffusion timescale for the entire outer cocoon.

{Since the kinetic energy of the outer cocoon originates from its internal energy, we estimate the initial post-breakout internal energy of the diffusion material as $E_{\rm oc,int0}\approx (3\theta_{\rm j}^2/2)^{1/3}m_{\rm oc,diff}v_{\rm oc,diff}^2/2$, where the prefactor $(3\theta_{\rm j}^2/2)^{1/3}$ accounts for adiabatic losses during the transition of the outer cocoon’s geometry from a cylindrical to a spherical form, corresponding to a change in volume from $\sim2\pi \theta_{\rm j}^2r_\star^3$ to $\sim4\pi r_\star^3/3$ shortly after breakout. Since the cocoon roughly starts with an initial radius of $r_\star$ and undergoes adiabatic cooling during expansion, with the internal energy evolving with radius as $E_{\rm oc,int}\propto 1/r_{\rm oc}$, the internal energy at time $t$ can be expressed as $E_{\rm oc,int}(t)\approx E_{\rm oc,int0}(r_\star/v_{\rm oc,diff}t)$. } The observed luminosity is simply $L_{\rm oc}\approx E_{\rm oc,int}/t$, which can be further expressed as\footnote{Since the outer cocoon may rapidly become spherical after breakout, we ignore the factor of $\theta^2/2$ that accounts for the angular correction to the luminosity, as included in Equation (39) of \cite{Piro2018}.}
\begin{equation}
\begin{split}
\label{equ:OuterCocoonCoolingEmission1}
    &L_{\rm oc}(t)\approx\left(\frac{3\theta_{\rm j}^2}{2}\right)^{4/3}\frac{M_{\rm ej}v_{\rm oc,min}r_\star}{2t_{\rm oc,diff}^2}\left(\frac{t}{t_{\rm oc,diff}}\right)^{-4/(s+2)} \\
    &\approx2.2\times10^{43}\,E_{\rm oc,52}^{2/3}\theta_{\rm j,-0.6}^{5/3}r_{\star,11.5}\left(\frac{t}{1\,{\rm day}}\right)^{-4/3}{\rm erg}\,{\rm s}^{-1},\  (s=1)
\end{split}
\end{equation}
where the second approximation is for $s=1$ following numerical simulations \citep[e.g.,][]{Nakar2017}. \cite{Srinivasaragavan2025}  revealed that the bolometric luminosities of the pre-SN bump in SN\,2025kg at $\sim1\,{\rm day}$ and $\sim2.5\,{\rm days}$ are $\sim4\times10^{43}{\rm erg}\,{\rm s}^{-1}$ and $\sim1\times10^{43}{\rm erg}\,{\rm s}^{-1}$, respectively, which can be roughly reproduced if the outer cocoon has properties of $\theta_{\rm j}\sim15^\circ$, $M_{\rm ej}\sim2\,M_\odot$, $E_{\rm oc}\sim2\times10^{52}{\rm erg}$ (or $v_{\rm oc,min}\sim0.3\,c$), and $r_\star\sim5\,R_\odot$.

\subsubsection{EP250108a from Inner Cocoon Cooling Emission} \label{sec:InnerCocoonModel}

Similar to the outer cocoon, most of the internal energy in the inner cocoon is expected to be converted into bulk kinetic energy. Motivated by numerical simulations suggesting that the relativistic inner cocoon has a relatively narrow distribution in proper velocity \citep[e.g.,][]{Gottlieb2016,Gottlieb2021}, we adopt a constant Lorentz factor $\gamma_{\rm ic}$ to describe its bulk motion. A typical opening angle of the inner cocoon shortly after the breakout is $\theta_{\rm ic}\sim30^\circ$ \citep[e.g.,][]{Nakar2017}. The optical depth of the visible inner cocoon material can be estimated as $\tau_{\rm ic}(>m_{\rm ic,diff})\sim \kappa m_{\rm ic,diff}/2\pi\theta_{\rm ic}^2 r_{\rm ic}^2\sim1$, where $r_{\rm ic}\sim2\gamma_{\rm ic}^2ct^2$. When all photons within the inner cocoon are able to diffuse out, i.e., $m_{\rm ic,diff}\sim E_{\rm ic}/\gamma_{\rm ic}c^2$, one can estimate the diffusion timescale for the entire inner cocoon as
\begin{equation}
    t_{\rm ic,diff}\approx\left(\frac{\kappa E_{\rm ic}}{8\pi\theta_{\rm ic}^2\gamma_{\rm ic}^5c^4}\right)^{1/2}\approx250\,E_{\rm ic,52}^{1/2}\theta_{\rm ic,-0.3}^{-1}\gamma_{\rm ic,0.7}^{-5/2}\,{\rm s},
\end{equation}
after which the emission is expected to decay rapidly.

The internal energy of the diffusion material can be expressed as $E_{\rm ic,int}\approx\gamma_{\rm ic}m_{\rm ic,diff}c^2\cdot(V_{\rm c}/V'_{\rm ic})^{1/3}\gamma_{\rm ic}$. The factor $(V_{\rm c}/V'_{\rm ic})^{1/3}\gamma_{\rm ic}$ accounts for the volume change, leading to the adiabatic loss, as the inner cocoon expands from a cylinder inside the progenitor star, with a total two-sided volume $V_{\rm c}\sim2\pi\theta_{\rm j}^2r_\star^3$, to a conical shape with a two-sided comoving volume $V'_{\rm ic}\sim4\pi\theta_{\rm ic}^2r_{\rm ic}^3/\gamma_{\rm ic}$. This internal energy is released over time $t$, implying an isotropic-equivalent luminosity of
\begin{equation}
\begin{split}
\label{equ:InnerCocoonCoolingEmission1}
    L_{\rm ic}&\approx\frac{2}{\theta_{\rm ic}^2}\frac{E_{\rm ic,int}}{t}= \frac{2^{8/3}\pi\theta_{\rm j}^{2/3}\gamma_{\rm ic}^{13/3}c^3r_\star}{\kappa\theta_{\rm ic}^{2/3}} \\
    & = 1.2\times10^{48}\theta_{\rm j,-0.6}^{2/3}\theta_{\rm ic,-0.3}^{-2/3}\gamma_{\rm ic,0.7}^{13/3}r_{\star,11.5}\,{\rm erg}\,{\rm s}^{-1},
\end{split}
\end{equation}
which is independent of $t$. The peak energy of the emission spectrum can be characterized by the blackbody temperature, specifically
\begin{equation}
\begin{split}
    E_{\rm ic,peak}&\approx4k_{\rm B}\mathcal{D}_{\rm on}T'_{\rm ic,eff} \\
    &\approx 0.36\,\theta_{\rm j,-0.6}^{1/6}\theta_{\rm ic,-0.3}^{-1/6}\gamma^{7/12}_{\rm ic,0.7}r^{1/4}_{\star,11.5}\left(\frac{t}{t_{\rm ic,diff}}\right)^{-1/2}{\rm keV}
\end{split}
\end{equation}
where the Doppler factor for the on-axis observer is $\mathcal{D}_{\rm on}\sim2\gamma_{\rm ic}$ and the effective temperature in the comoving frame is $T'_{\rm ic,eff}=(E'_{\rm ic,int}/aV'_{\rm ic})^{1/4}\approx (L_{\rm ic}/4\pi\sigma_{\rm SB}\mathcal{D}_{\rm on}^2r_{\rm ic}^2)^{1/4}$ with $a=4\sigma_{\rm SB}/c$ being the radiation constant. For an observer with a viewing angle of $\theta_{\rm v}$, the observed luminosity, diffusion timescale, and peak energy can be roughly estimated by $L^{\rm obs}_{\rm ic}\approx\mathcal{D}^2L_{\rm ic}/\mathcal{D}_{\rm on}^2$, $t_{\rm ic,diff}^{\rm obs}\approx \mathcal{D}_{\rm on}t_{\rm ic,diff}/\mathcal{D}$, and $E_{\rm ic,peak}^{\rm obs}=\mathcal{D}E_{\rm ic,peak}/\mathcal{D}_{\rm on}$, respectively, where $\mathcal{D}\approx\gamma^{-1}_{\rm ic}\cos^{-1}[\max(\theta_{\rm v}-\theta_{\rm ic},0)]$ represents the Doppler factor.

The X-ray prompt emission of EP250108a exhibits several unique properties \citep{Li2025}, including a duration of $T_{90}\sim960\,{\rm s}$, a peak energy of $<1.8\,{\rm keV}$, which may even fall below the lower bound of WXF's energy range (i.e., $0.5\,{\rm keV}$), an isotropic X-ray luminosity of $(1-4.4)\times10^{46}{\rm erg}\,{\rm s}^{-1}$, and an isotropic-equivalent X-ray energy of $(5-30)\times10^{48}{\rm erg}$. If the inner cocoon is viewed on-axis ($\theta_{\rm v}\lesssim\theta_{\rm ic}$), its cooling emission would be significantly brighter and of much shorter duration compared to that observed in EP250108a. However, the observed properties of EP250108a could be explained by the inner cocoon cooling emission viewed slightly off-axis. For example, we then adopt $\gamma_{\rm ic}\sim5$, $\theta_{\rm ic}\sim30^\circ$, and $\theta_{\rm v}\sim45^\circ$, as well as similar parameters to those inferred for the outer cocoon in Section \ref{sec:OuterCocoonModel}, including $E_{\rm ic}\sim1\times10^{52}{\rm erg}$, $r_\star\sim5\,R_\odot$, and $\theta_{\rm c}\sim15^\circ$ to infer the observed properties for the inner cocoon cooling emission. One can derive a duration of $t_{\rm ic,diff}^{\rm obs}\sim800(1+z)\,{\rm s}$, an isotropic X-ray luminosity of $L_{\rm ic,X}^{\rm obs}\sim 4\times10^{46}\eta_{\rm X,-0.5}\,{\rm erg}\,{\rm s}^{-1}$, an isotropic-equivalent X-ray energy of $E_{\rm ic,X}^{\rm obs}\sim L_{\rm ic,X}^{\rm obs}t_{\rm ic,diff}^{\rm obs}\sim 3\times10^{49}\eta_{\rm X,-0.5}\,{\rm erg}$, and a peak energy of $E_{\rm ic,peak}^{\rm obs}\sim0.2(1+z)\,{\rm keV}$, expected to appear around the diffusion time, where $\eta_{\rm X}$ is the X-ray-to-total luminosity ratio. The inferred parameters lie within the range of the observed values, suggesting that EP250108a likely originates from off-axis inner cocoon cooling emission.

\section{Model and Lightcurve Fits for SN\,2025kg}

\subsection{Outer Cocoon Cooling Emission}

After the diffusion timescale, i.e., the time $t>t_{\rm oc,diff}$, all thermal photons trapped in the outer cocoon are able to diffuse out. As a result, the entire outer cocoon continues to cool via radiation and adiabatic expansion, leading to an exponential decline in luminosity \citep{Piro2021}. Thus, the luminosity of the cooling emission from the outer cocoon, given by Equation (\ref{equ:OuterCocoonCoolingEmission1}), can be more precisely expressed as
\begin{equation}
\begin{split}
\label{equ:OuterCocoonCoolingEmission2}
    L_{\rm oc}(t)\approx&\left(\frac{3\theta_{\rm j}^2}{2}\right)^{4/3}\frac{M_{\rm ej}v_{\rm oc,min}r_\star}{2t_{\rm oc,diff}^2} \\
    &\times\begin{cases}
    (t/t_{\rm oc,diff})^{-4/(s+2)}, & \text{if}\;t\leq t_{\rm oc,diff}, \\
    e^{-(t^2/t_{\rm oc,diff}^2-1)/2}, & \text{if}\;t> t_{\rm oc,diff}.
\end{cases}
\end{split}
\end{equation}
The photospheric velocity can be derived by $\tau_{\rm oc}(>v_{\rm oc,ph})\sim1$, which gives $v_{\rm oc,ph}\approx v_{\rm oc,min}(t/t_{\tau\sim1})^{-2/(s+3)}$, where $t_{\tau\sim1}\approx(3\theta_{\rm j}^2\kappa M_{\rm ej}/8\pi v_{\rm oc,min}^2)^{1/2}=2.7\,E_{\rm oc,52}^{-1/2}\theta_{\rm j,-0.6}(M_{\rm ej}/2\,M_\odot)\,{\rm days}$ is the characteristic timescale at which the entire outer cocoon becomes optically thin. The spectral luminosity for a given observed frequency $\nu$ can be calculated as
\begin{equation}
\label{equ:CocooonCoolingEmission}
    L_{\rm \nu,oc}(\nu,t)\approx4\pi v_{\rm oc,ph}^2t^2\mathcal{D}\cdot \pi B_\nu[\nu(1+z)/\mathcal{D},T'_{\rm oc,eff}]
\end{equation}
where $T'_{\rm oc,eff}(t)\approx(L_{\rm oc}/4\pi\sigma_{\rm SB}\mathcal{D}^2v_{\rm oc,ph}^2t^2)^{1/4}$ is the comoving-frame effective temperature, $B_\nu$ is the Planck function, $\mathcal{D}\approx1/(1-v_{\rm oc,ph}/c)$ is the Doppler factor, and $\sigma_{\rm SB}$ is the Stefan-Boltzmann constant. Then, the observed flux is $F_{\rm \nu,oc}=(1+z)L_{\rm \nu,oc}/4\pi D_{\rm L}^2$. 

\subsection{SN Emission Driven by a Magnetar Central Engine and Radioactive Decay}
\label{sec:SNEmissionModel}

The energy conservation for the internal energy $E_{\rm ej,int}$ of the SN ejecta is described by
\begin{equation}
    \frac{dE_{\rm ej,int}}{dt}=L_{\rm sd}(t) + L_{\rm rad}(t) - L_{\rm ej}(t)-p_{\rm ej}\frac{dV_{\rm ej}}{dt},
\end{equation}
where $p_{\rm ej}=3E_{\rm ej,int}/4\pi r_{\rm ej}^3$ is the pressure related to the internal energy, and the term $p_{\rm ej}dV_{\rm ej}/dt=4\pi r_{\rm ej}^2p_{\rm ej}v_{\rm ej}$, with $v_{\rm ej}=dr_{\rm ej}/dt$ being the velocity, represents the rate of energy loss due to adiabatic expansion. The radioactivity luminosity is $L_{\rm rad}(t)=(1-e^{-\tau_\gamma})L_{\rm Ni}(t)$, where $\tau_\gamma\approx3\kappa_\gamma M_{\rm ej}/4\pi r_{\rm ej}^2$ represents the optical depth of the ejecta to gamma-ray photons produced by radioactive decay, with $\kappa_\gamma$ being the gamma-ray opacity \citep[e.g.,][Zhu 2025 in prep]{Wheeler2015}. The adiabatic loss is converted into the kinetic energy of the ejecta, and hence the dynamical evolution is governed by 
\begin{equation}
\label{equ:SNEjectaVelocity}
    dv_{\rm ej}/dt=4\pi r_{\rm ej}^2p_{\rm ej}/M_{\rm ej}.
\end{equation}
Following \cite{Kasen2010}, one can express the bolometric luminosity empirically as $L_{\rm ej}(t) \approx cE_{\rm ej,int}(1-e^{-\tau_{\rm ej}})/r_{\rm ej}\tau_{\rm ej}$
with the total optical depth given by $\tau_{\rm ej}\approx3\kappa M_{\rm ej}/4\pi r_{\rm ej}^2$. 

Observationally, the measurable parameters include the photosphere radius, for which we assume $r_{\rm ej,ph}\simeq r_{\rm ej}$ based on the standard approximation in the literature. Furthermore, motivated by observations and lightcurve fittings for SNe Ic-BL {\citep[e.g.,][]{Inserra2013,Nicholl2017}}, we introduce a floor temperature {caused by recombination of the ejecta and/or the influence of the central magnetar.} Then, the final effective temperature and photosphere radius can be given by $T_{\rm ej,eff}\approx\max[({L_{\rm ej}}/{4\pi\sigma_{\rm SB}r_{\rm ej}^2})^{1/4},T_{\rm floor}]$ and $r_{\rm ej,ph}\approx\min[r_{\rm ej},(L_{\rm ej}/4\pi\sigma_{\rm SB}T_{\rm floor}^4)^{1/2}]$. One can set $\mathcal{D}\approx1$ and replace $v_{\rm oc,ph}t$ and $T_{\rm oc,eff}$ in Equation (\ref{equ:CocooonCoolingEmission}) with $r_{\rm ej,ph}$ and $T_{\rm ej,eff}$, respectively, to calculate the spectral luminosity from the SN emission.

\subsection{Lightcurve Fits for SN\,2025kg and Results} \label{sec:BestFittings}

\begin{figure}
\centering
\includegraphics[width=0.95\linewidth, trim = 80 30 105 60, clip]{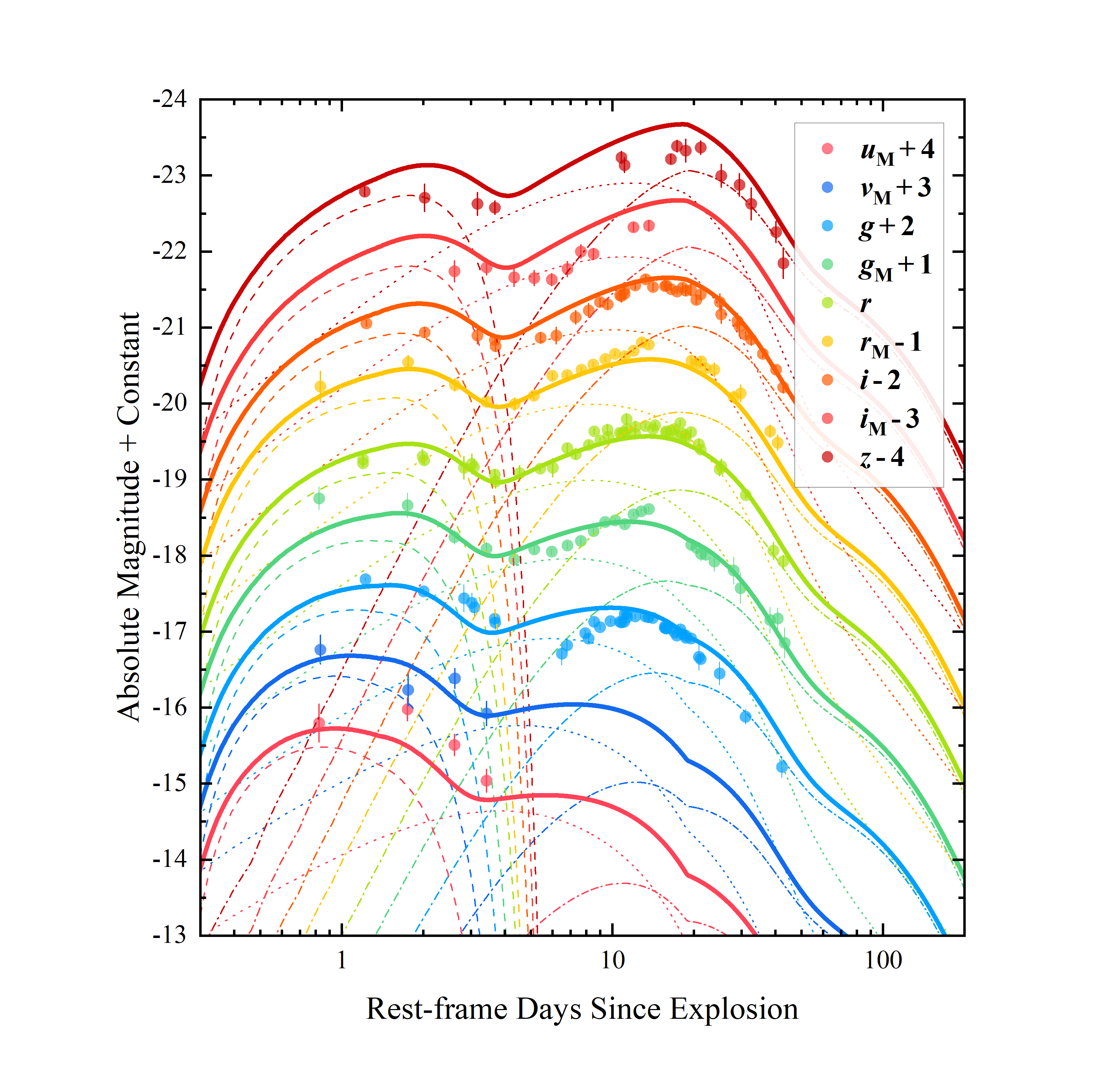}
\caption{Multi-band lightcurve fits for the data of SN\,2025kg \citep{Li2025}. The solid lines represent the best fits for the multi-band lightcurves. {The contributions from the outer cooling emission are marked by the dashed lines, while the dotted and dashed-dotted lines represent the SN emission input from the magnetar and $^{56}$Ni, respectively.} The colors corresponding to different bands are shown in the top-right legend. Note that the subscript ``M'' denotes the Mephisto filter system, which differs from the Sloan $griz$ system. }
\label{fig:OpticalLightcurve}
\end{figure} 

Adopting a Markov Chain Monte Carlo method with the \texttt{emcee} package \citep{ForemanMackey2013}, we employ the outer cocoon cooling emission model to fit the pre-SN bump of SN\,2025kg, while attributing the main peak to the magnetar spin-down engine and the late-time SN lightcurve to radioactive decay. 

When fitting the lightcurve of SN\,2025kg, we adopt $\kappa=0.1\,{\rm cm}^2{\rm g}^{-1}$ and $\theta_{\rm j}=15^\circ$, as defined based on the assumptions described above. Due to the lack of observations after $\sim40\,{\rm days}$, the $^{56}$Ni decay tail in the lightcurve of SN\,2025kg was not captured, making it hard to constrain $M_{\rm Ni}$ and $\kappa_\gamma$ directly. We therefore assume a fiducial value of $f_{\rm Ni}=0.1$, corresponding to the upper limit of typical values reported in the literature \citep[e.g.,][Zhu 2025 in prep]{Lv2018,Taddia2019}, and adopt $\kappa_\gamma\approx0.4\,{\rm cm}^2{\rm g}^{-1}$, which is the median value inferred from fitting results for the SN Ic-BL population (e.g., Zhu 2025 in prep). In addition, we assume an initial explosion energy of $E_{\rm SN}\sim10^{51}{\rm erg}$, a typical value for core-collapse SNe, which yields an initial ejecta velocity of $v_{\rm ej0}=\sqrt{2E_{\rm SN}/M_{\rm ej}}$ for Equation (\ref{equ:SNEjectaVelocity}). Similar to \cite{Li2025}, we adopt a standard Galactic reddening law with $R_{V}=3.1$ \citep{Cardelli1989}, and consider only the Galactic extinction along the line of sight to EP250108a/SN\,2025kg, for which $E(B-V)=0.015\,{\rm mag}$ \citep{Schlafly2011}. Finally, there are six free parameters {(see Appendix \ref{app_sec:PriorPosterior} for the priors used and the resulting posteriors of the fitting parameters)}, with constrained posteriors (with $68\%$ credible intervals) of $P_{\rm i}=1.65^{+0.02}_{-0.02}\,{\rm ms}$, $B_{\rm p}=2.01^{+0.01}_{-0.01}\times10^{15}{\rm G}$, $M_{\rm ej}=2.53^{+0.04}_{-0.03}\,M_\odot$, $T_{\rm floor}=6.72^{+0.14}_{-0.13}\times10^{3}{\rm K}$, $v_{\rm oc,min}=0.32^{+0.01}_{-0.01}\,{c}$ (corresponding to an energy of $E_{\rm oc}\sim2.2\times10^{52}{\rm erg}$), and $r_\star=4.64^{+0.33}_{-0.31}\,R_\odot$, which are basically consistent with our earlier estimates in Section \ref{sec:AnalyticalEstimate}.

The best-fitting multi-band lightcurves are shown in Figure \ref{fig:OpticalLightcurve} {(see Figure \ref{fig:BolometricEmission} in Appendix \ref{app_sec:BolometricLightcurve} for corresponding bolometric lightcurve)}. It shows that the multi-band lightcurves of the pre-SN bump before $\sim3\,{\rm days}$ and the main SN emission can be well explained by the outer cocoon cooling emission and the combined contributions from a magnetar central engine and radioactive decay, respectively. Most of ($\gtrsim98\%$) the magnetar rotational energy, $E_{\rm rot,i}\sim7\times10^{51}(P_{\rm i}/1.7\,{\rm ms})^{-2}{\rm erg}$, would be converted into the kinetic energy of the ejecta. The resulting total kinetic energy $8\times10^{51}{\rm erg}$ is considerably higher than that of normal core-collapse SNe, thereby producing the characteristic broad-line features in the spectra. From our lightcurve fittings in Figure \ref{fig:OpticalLightcurve}, the remaining fraction of the magnetar energy, though small, would contribute a significant portion ($\gtrsim50\%$) of the bolometric peak luminosity of SN\,2025kg around $\sim14\,{\rm days}$, as well as the majority of the pre-peak ultraviolet and optical emission. In contrast, the post-peak optical and infrared emission is primarily powered by radioactive decay.

\section{Progenitor of EP250108a/SN\,2025kg}

\subsection{A Close Helium-star Binary Origin of SN\,2025kg}

SN\,2025kg, classified as a SN Ic-BL, could originate from a low-mass stripped-envelope star with an ejecta mass of $M_{\rm ej}\sim2.5\,M_\odot$ and hence a pre-SN progenitor mass of $M_{\rm He}\sim4\,M_\odot$ if the remnant is an NS. The pre-SN progenitor's radius is inferred to be $\sim5\,R_\odot$, significantly larger than that of carbon-oxygen WR stars, which are commonly considered the progenitors of SNe Ic-BL and LGRBs and are usually very compact with radii of $\lesssim1\,R_\odot$ \citep[e.g.,][]{AguileraDena2018}. However, a radius of $\sim5\,R_\odot$ roughly corresponds to that of a $\sim4\,M_\odot$ pre-SN helium star at solar metallicity \citep[e.g.,][]{Yoon2010,Woosley2019}, suggesting that the helium envelope might not be fully stripped before the explosion of SN\,2025kg. The presence of a low-mass extended helium envelope surrounding the carbon-oxygen core is also supported by the $\sim1$ and $2\mu{\rm m}$  He \textsc{i} absorption features identified in the JWST spectrum \citep{Rastinejad2025}.

Furthermore, forming a magnetar with an initial spin period of $\sim1.7\,{\rm ms}$ requires a rapidly rotating pre-SN progenitor. The host of EP250108a/SN\,2025kg is metal-rich, with a metallicity comparable to solar, i.e., $Z \sim Z_\odot$ \citep{Li2025}. A single progenitor star with such metallicity is unlikely to retain sufficient angular momentum throughout its evolution to form a rapidly rotating magnetar, as the magnetic torques \citep{Spruit2002,Fuller2019} can efficiently enhance angular momentum loss via stellar winds, thereby slowing the stellar core. Furthermore, this high-metallicity environment poses a serious challenge to one of the most popular and widely accepted progenitor scenarios for SNe Ic-BL and LGRBs, i.e., the CHE channel, which is predicted to occur only at very low metallicities of $Z < 0.3\,Z_\odot$ \citep[e.g.,][]{Woosley2006,Yoon2006}. 

We then test whether SN\,2025kg could originate from a close binary system produced through the isolated binary evolution channel, either via common-envelope evolution (CEE) or stable mass transfer (SMT), in which the helium-star progenitor can be tidally spun up by its companion if the orbital period is $P_{\rm orb}\lesssim2\,{\rm days}$ and potentially leave behind a rapidly rotating magnetar upon core collapse \citep{Fuller2022,Hu2023}. In such a binary, the helium star must stay within its Roche lobe, setting an upper limit on its radius. Assuming a mass ratio of $q\sim1$ and $P_{\rm orb}\sim2\,{\rm days}$, the Roche lobe radius can be approximated using the \cite{Eggleton1983} formula: $R_{\rm L}=0.49q^{2/3}a/[0.6q^{2/3}+\ln(1+q^{1/3})]\sim0.38a\sim5\,R_\odot$, where $a\sim (GM_{\rm He}P_{\rm orb}^2/2\pi^2)^{1/3}\sim13\,(M_{\rm He}/4\,M_\odot)^{1/3}(P_{\rm orb}/2\,{\rm day})^{2/3}\,R_\odot$ is the binary separation. Therefore, the progenitor of SN\,2025kg was likely a $\sim4\,M_\odot$ helium star in a close binary system with a pre-SN orbital period of $\sim2\,{\rm days}$, which probably filled its Roche lobe and experienced Case BB/BC mass transfer before the explosion.

\subsection{Weak Stellar Wind Revealed by Missing Cocoon Afterglow } \label{sec:Afterglows}

\begin{figure}
\centering
\includegraphics[width=1\linewidth, trim = 80 30 55 55, clip]{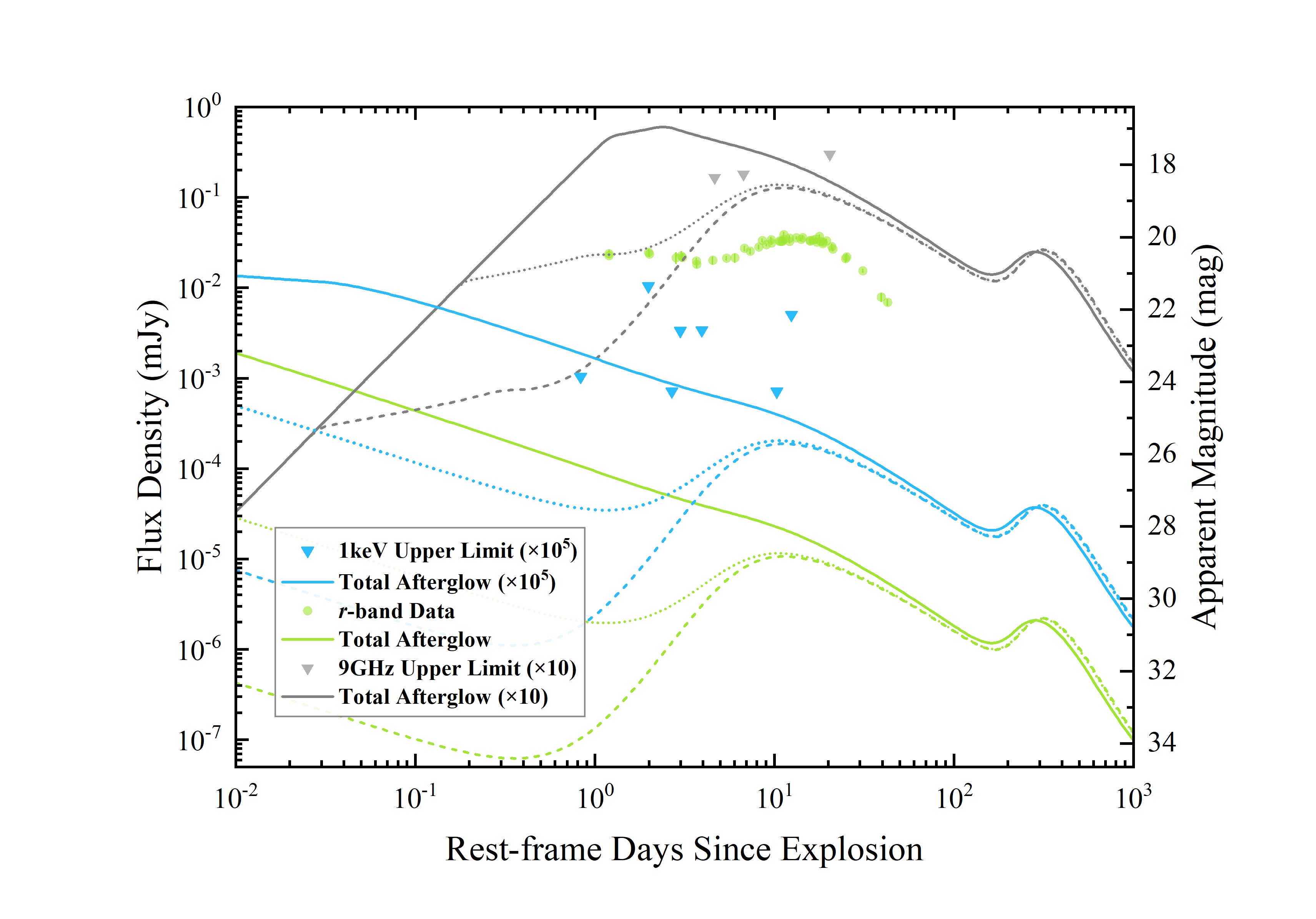}
\caption{1\,{\rm keV} (blue), $r$-band (green), and $9\,{\rm GHz}$ (gray) simulated afterglow lightcurves. The solid, dotted, and dashed lines represent $A_\star=10^{-1}$, $10^{-2}$, and $10^{-3}$, respectively. The data are collected from \citet{Li2025}.}
\label{fig:AfterglowLightcurve}
\end{figure} 

{Both the jet and inner cocoon can independently generate observable afterglow emissions from radio to X-ray, while the afterglow from the transrelativistic outer cocoon would appear much later and be faint, making it difficult to detect.} In order to simulate the afterglows {(see Appendix \ref{app_sec:AfterglowModel} for details of our afterglow model)}, we adopt representative microphysical parameters with an electron spectral index $p=2.3$, and fractions of energy in electrons and magnetic fields of $\epsilon_e=0.1$ and $\epsilon_B=10^{-4}$, respectively. The circumburst environment is modeled with a radial density profile given by $n(r)= 3\times10^{35} A_\star r^{-2}{\rm cm}^{-1} + n_{\rm ISM}$, where $A_\star=(\dot{M}/10^{-5}M_\odot\,{\rm yr}^{-1})(v_{\rm w}/10^3{\rm km}\,{\rm s}^{-1})^{-1}$ is the wind density parameter normalized by the mass-loss rate and wind velocity, and the interstellar density at large radii is set to $n_{\rm ISM}=0.1\,{\rm cm}^{-3}$. We adopt $\theta_{\rm v}\sim45^\circ$ and the inner cocoon parameters of $E_{\rm ic}=10^{52}{\rm erg}$, $\theta_{\rm ic}=30^\circ$, and $\gamma_{\rm ic}=5$ following Section \ref{sec:InnerCocoonModel}. The jet parameters used include a jet energy $E_{\rm j}=10^{52}{\rm erg}$, $\theta_{\rm j}=15^\circ$, and a Lorentz factor $\gamma_{\rm j}=300$. Figure \ref{fig:AfterglowLightcurve} shows the simulated multi-band afterglow lightcurves for $A_\star$ values between $10^{-3}$ and $10^{-1}$. The afterglow emission is attributed to the inner cocoon at times before $\sim100\,{\rm days}$ and to the jet at later times. An afterglow below the observational limits requires $A_\star\lesssim10^{-2}$; a higher $\epsilon_B>10^{-4}$ requires an even lower $A_\star$, implying a weak wind for the progenitor of EP250108a/SN\,2025kg. This inferred $A_\star$ is at least an order of magnitude lower than that of Galactic WR stars (i.e., $Z\sim Z_\odot$), where $A_\star$ is always larger than $0.12$ \citep{Nugis2000}. Such a weak wind is theoretically consistent with the mass-loss properties of low-mass helium stars \citep{Vink2017}, further supporting the inferred radius and mass of the progenitor of EP250108a/SN\,2025kg discussed above.

\subsection{MS Companion as Source of Late-time Broad H$\alpha$ Line}

\cite{Rastinejad2025} reported a broad H$\alpha$ feature at $\sim42.5\,{\rm days}$, absent at other epochs, and attributed to interaction between the ejecta and circumstellar material (CSM). The velocity of the hydrogen-rich material can be estimated as $v_{\rm H\alpha}\sim \Delta\lambda_{\rm FWHM}c/2\lambda_{\rm H\alpha}\sim3540\,(\Delta\lambda_{\rm FWHM}/150\text{\AA})(\lambda_{\rm H\alpha}/6355\text{\AA})^{-1}{\rm km}\,{\rm s}^{-1}$, where $\Delta\lambda_{\rm FWHM}\sim150\text{\AA}$ is an estimate for the H$\alpha$ line's full width at half maximum based on Figure 3 of \cite{Rastinejad2025}. This H$\alpha$ velocity is comparable to that of the Si \textsc{ii} line, i.e., $v_{\rm Si}\sim4000\,{\rm km}\,{\rm s}^{-1}$, observed at the same epoch, suggesting that the H$\alpha$ may originate from the low-velocity inner ejecta rather than ejecta-CSM interaction. Thus, we suggest this H$\alpha$ may plausibly be attributed to material partially stripped from a MS companion during the explosion.

\cite{Zhu2024} proposed a ``magnetar--star binary engine'' model for superluminous SNe (SLSNe), in which a newborn magnetar can drive the SLSN emission and simultaneously evaporate its MS companion by the powerful magnetar wind within the enclosure of the SN ejecta. These evaporated stellar materials would be accelerated by the magnetar wind and located in the inner part of the ejecta, resulting in late-time broad H$\alpha$ \citep{Yan2015,Yan2017} after the ejecta becomes optically thin. We suggest that this model may also apply to SNe Ic-BL, provided that their progenitors are close binary systems composed of a helium star and an MS companion. However, the observed H$\alpha$ luminosity and appearance time in SNe Ic-BL are expected to be markedly different from those in SLSNe due to the typically much higher magnetic fields and, hence, shorter spin-down timescale of the magnetars in SNe Ic-BL \citep[e.g.,][]{Metzger2015,Yu2017,Liu2022}. The mass of evaporated hydrogen-rich material is determined by the $M_{\rm ev}\propto L_{\rm sd}t_{\rm sd}\propto E_{\rm rot,i}$ \citep{Zhu2024}, implying a comparable hydrogen mass in SLSNe and SNe Ic-BL. However, producing observable H$\alpha$ emission requires this material to be excited by the central engine, with luminosity $L_{\rm H\alpha}\propto L_{\rm sd}+L_{\rm rad}$. In SLSNe, the late-time spin-down power remains significant, while in SNe Ic-BL, the energy source at late phases should be dominated by radioactive decay, with $L_{\rm rad}$ being one to two orders of magnitude lower than $L_{\rm sd}$ in SLSNe. This implies that H$\alpha$ emission in SNe Ic-BL should be considerably fainter and easily missed. Compared to SLSNe, more magnetar rotational energy in SNe Ic-BL is converted into kinetic energy. As a result, the ejecta expands faster, with less energy being converted into thermal energy, leading to lower opacity at late times. Consequently, SNe Ic-BL become transparent much earlier, with the appearance timescale $t_{\tau\sim1}\sim\sqrt{3\kappa M_{\rm ej}^2/8\pi E_{\rm ej,kin}}\sim 36\,\kappa_{-1.3}^{1/2}E_{\rm ej,kin,52}^{-1/2}(M_{\rm ej}/2\,M_\odot)\,{\rm days}$, which is comparable to the observed epoch of H$\alpha$ line in SN\,2025kg, where $\kappa\sim0.05\,{\rm cm}\,{\rm g}^{-1}$ is the opacity appropriate for late-time low-temperature carbon-oxygen material \citep[e.g.,][]{Piro2014}.  

Thus, one can conclude the late-time broad H$\alpha$ line observed in SN\,2025kg by \citet{Rastinejad2025} could plausibly originate from the evaporation of a MS companion by the newborn magnetar. Alternatively, \cite{Moriya2015} suggested that late-time broad H$\alpha$ emission could be produced by hydrogen material stripped from a MS companion star during the SN explosion \citep{Hirai2018}. However, we note that simulations remain highly uncertain regarding the efficiency and viability of such stripping \citep[e.g.,][]{Liu2015}.

\section{Conclusions}

In this {\em Letter}, we have proposed that the prompt emission of EP250108a may originate from the off-axis cooling emission of a mildly relativistic inner cocoon, with a viewing angle of $\sim45^\circ$, while the cooling emission of the outer cocoon accounts for the early bump observed in SN\,2025kg. Both the outer and inner cocoons are constrained to have an energy of $\sim(1-2)\times10^{52}{\rm erg}$, and the progenitor's radius is estimated to be $\sim5\,R_\odot$. 

{The energy of the jet-cocoon system may originate from magnetic dissipation due to differential rotation \citep{Kluzniak1998} or accretion \citep{ZhangD2010} of the newborn magnetar. Subsequently, } a magnetar with $P_{\rm i}\sim1.7\,{\rm ms}$ and $B_{\rm p}\sim2\times10^{15}{\rm G}$ may have been left behind, whose magnetic dipole spin-down energy injection powered the extremely luminous peak of SN\,2025kg. The progenitor mass is constrained to be $\sim4\,M_\odot$, suggesting that the progenitor was likely a helium star with an expanded envelope of $5\,R_\odot$, supported by the observed He \textsc{i} absorption lines \citep{Rastinejad2025} and the weak wind inferred for the pre-SN progenitor, as evidenced by the absence of a detectable inner cocoon afterglow. The late-time broad H$\alpha$ line \citep{Rastinejad2025} indicates that the helium star existed in a close binary system with an MS companion, and it may be produced by hydrogen-rich MS stellar material evaporated by the powerful magnetar wind, which became visible once the SN ejecta became optically thin. The pre-SN orbital period is likely $\sim2\,{\rm days}$, allowing the helium star to be tidally spun up by the MS companion and naturally form a rapidly rotating magnetar. Given the high metallicity of this event, which may make CHE difficult to occur, we have suggested that EP250108a/SN\,2025kg most likely originated through the isolated binary evolution channel.

We note that \cite{Li2025} also suggested a magnetar-powered scenario for SN\,2025kg but with a different set of magnetar parameters, i.e., $P_{\rm i}\sim14.5\,{\rm ms}$ and $B_{\rm p}\sim2.6\times10^{14}{\rm G}$. This is because \citet{Li2025} suggested that EP250108a may originate from an on-axis weak jet and therefore requires a relatively slowly rotating magnetar. In contrast, we propose that EP250108a may be powered by the inner cocoon of a typical-energy GRB jet viewed off-axis. The inferred magnetar in SN\,2025kg is a millisecond magnetar, which is more consistent with those proposed in standard LGRB models \citep{Usov1992,Dai1998,Zhang2001,Zhang2018}.

\section*{Data AVAILABILITY}

All data presented in this {\em Letter} are obtained from \citet{Li2025}. The analysis codes are available upon reasonable request from the corresponding authors.

\section*{Acknowledgements}

{We note that a related study by \cite{Gottlieb2025} appeared after our preprint was posted, which also attributed the early bump to cocoon cooling emission but invoked additional black-hole–accretion–disk–driven outflows.} We thank {the anonymous referee for valuable suggestions,} Wen-Xiong Li, Liang-Duan Liu, Wenbin Lu, Ilya Mandel, Guang-Lei Wu, and Yun-Wei Yu for their helpful comments and suggestions. JHZ’s work is supported by the National Natural Science Foundation of China (grant Nos. 124B2057).

\bibliographystyle{mnras}
\bibliography{ms} 

\appendix

\section{Schematic Picture} \label{app_sec:Illustration}

\begin{figure*}
\centering
\includegraphics[width=1\linewidth]{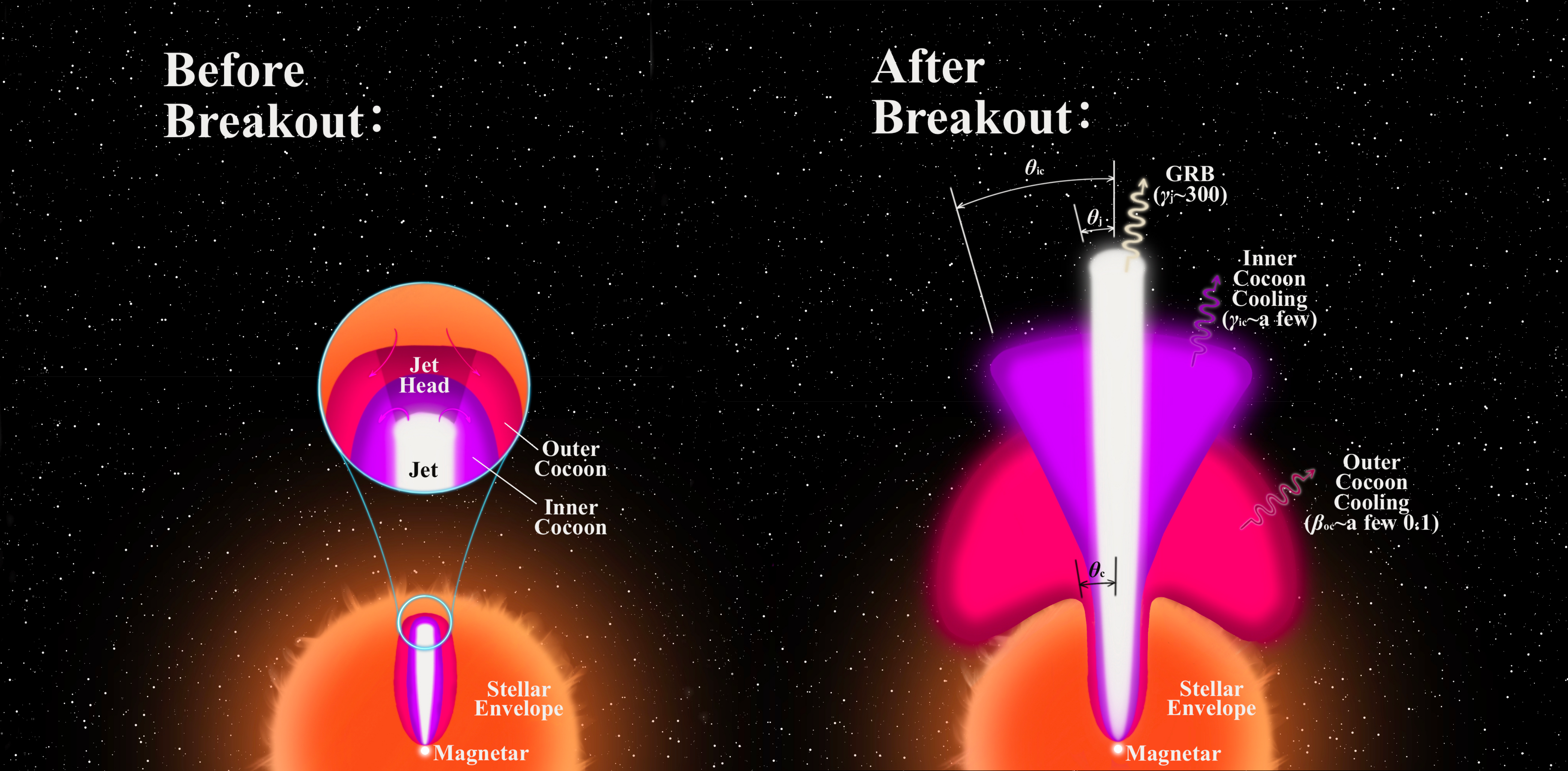}
\caption{{Schematic illustration of the jet structure before the breakout and the resulting emission components after the breakout.  }}
\label{fig:Illustration}
\end{figure*}

{Figure \ref{fig:Illustration} illustrates the jet structure before the breakout and the resulting emission components after the breakout. }

\section{Definitions of Frequently Used Variables}

{In Table \ref{tab:VariableDefinition}, we summarize the definitions of frequently used variables.}

\begin{table}
    \centering
    \caption{{Definitions of Frequently Used Variables}}
    \begin{tabular}{|l|p{6.75cm}|}
    \hline
    \hline
    Variable  &  \multicolumn{1}{c}{Definition} \\
    \hline
    $\theta_{\rm v}$ & Viewing angle measured from the jet axis \\
    $z$ & Redshift; the value for EP250108a/SN\,2025kg is $z=0.176$ \\
    $D_{\rm L}$ & Luminosity distance; the value for EP250108a/SN\,2025kg is $D_{\rm L}=881\,{\rm Mpc}$ \\
    $E(B-V)$ & Extinction; the value is $E(B-V)=0.015\,{\rm mag}$, which only includes Galactic extinction \\
    \hline
    \multicolumn{2}{c}{Magnetar-powered SN Explosion Emission} \\
    \hline
    $P_{\rm i}$ & Initial spin period of the magnetar \\
    $B_{\rm p}$ & Polar strength of the dipolar magnetic field \\
    $L_{\rm sd,i}$ & Initial spin-down luminosity, which is expressed as $L_{\rm sd,i}\simeq10^{47}B_{\rm p,14}^2P_{\rm i,-3}^{-4}{\rm erg}\,{\rm s}^{-1}$ \\
    $t_{\rm sd}$ & Spin-down timescale, which is $t_{\rm sd}\simeq2\times10^5B_{\rm p,14}^{-2}P_{\rm i,-3}^{2}{\rm s}$ \\
    $M_{\rm ej}$ & SN ejecta mass \\
    $M_{\rm Ni}$ & $^{56}$Ni mass \\
    $f_{\rm Ni}$ & Fraction of $^{56}$Ni in the ejecta mass, i.e., $f_{\rm Ni}=M_{\rm Ni}/M_{\rm ej}$; the value we set is $f_{\rm Ni}=0.1$ \\
    $\kappa$ & Gray opacity; the value we set is $\kappa\approx0.1\,{\rm cm}^2{\rm g}^{-1}$ \\
    $\kappa_\gamma$ & Opacity for gamma-ray photons; the value we set is $\kappa_\gamma \approx 0.4\,{\rm cm}^2{\rm g}^{-1}$\\
    $T_{\rm floor}$ & Floor temperature caused by recombination of the ejecta and/or the influence of the central magnetar, typically in $\sim4000-7000\,{\rm K}$ \citep[e.g.,][Zhu 2025 in prep]{Inserra2013,Nicholl2017} \\
    $E_{\rm SN}$ & SN explosion energy produced by the neutrino mechanism of core collapse, typically set to $E_{\rm SN}\sim10^{51}{\rm erg}$ \\
    $E_{\rm rot}$ & Rotational energy of the magnetar, which is $E_{\rm rot}=L_{\rm sd,i}t_{\rm sd}$ \\
    $r_{\rm ej,ph}$ & Photosphere radius of the SN ejecta \\
    $T_{\rm oc,eff}$ & Effective temperature of the SN ejecta\\
    \hline
    \multicolumn{2}{c}{Cocoon Cooling Emissions} \\
    \hline
    $r_\star$ & Radius of the progenitor \\
    $s$ & Power-law index describing the steepness of the outer cocoon’s velocity distribution; the value we set is $s=1$ \\
    $v_{\rm oc,min}$ & Minimum velocity of the outer cocoon \\
    $\gamma_{\rm ic}$ & Lorentz factor of the inner cocoon\\
    $\theta_{\rm j}$ & Jet half-opening angle, also serving as the half-opening angle of the outer and inner cocoons before breakout inside the progenitor; the value we set is $\theta_{\rm j}=15^\circ$, which is reasonable for jets propagating within a $\sim4\,M_\odot$ helium star progenitor \\
    $\theta_{\rm ic}$ & Half-opening angle of the inner cocoon after breakout; the value we set is $\theta_{\rm ic}=30^\circ$ \citep[e.g.,][]{Nakar2017} \\
    $M_{\rm oc}$ & Mass of the outer cocoon, expressed as $M_{\rm oc}\sim3\theta_{\rm j}^2M_{\rm ej}/2$  \\
    $E_{\rm oc}$ & Explosion energy of the outer cocoon, which can be estimated as $E_{\rm oc}\approx M_{\rm oc}v_{\rm oc,min}^2/2$  \\
    $E_{\rm ic}$ & Explosion energy of the inner cocoon \\
    $r_{\rm oc,ph}$ & Photosphere radius of the outer cocoon cooling emission \\
    $T_{\rm oc,eff}$ & Effective temperature of the outer cocoon cooling emission\\
    $E_{\rm ic,peak}$ & Peak energy of the inner cocoon cooling emission \\
    \hline
    \multicolumn{2}{c}{Afterglow Emissions} \\
    \hline
    $A_\star$ & Wind density parameter normalized by the mass-loss $\dot{M}$ and wind velocity $v_{\rm w}$, which is expressed as $A_\star=(\dot{M}/10^{-5}M_\odot\,{\rm yr}^{-1})(v_{\rm w}/10^3{\rm km}\,{\rm s}^{-1})^{-1}$ \\
    $n_{\rm ISM}$ & Interstellar density; the value is set to $n_{\rm ISM} = 0.1\,{\rm cm}^{-3}$ \\
    $\gamma_{\rm j}$ & Lorentz factor of the jet; the value is set to $\gamma_{\rm j}=300$ \\
    $E_{\rm j}$ & Jet energy; the value is set to $E_{\rm j}=10^{52}{\rm erg}$\\
    $p$ & Electron spectral index \\
    $\epsilon_e$ & Fraction of energy in electrons; the value we set is $\epsilon_e=0.1$ \\
    $\epsilon_B$ & Fraction of energy in magnetic fields; the value we set is $\epsilon_B = 10^{-4}$ \\
    \hline
    \end{tabular}
    \label{tab:VariableDefinition}
\end{table}

\section{Prior and Posterior Distributions of Fitting Parameters} \label{app_sec:PriorPosterior}

\begin{figure*}
\centering
\includegraphics[width=1\linewidth]{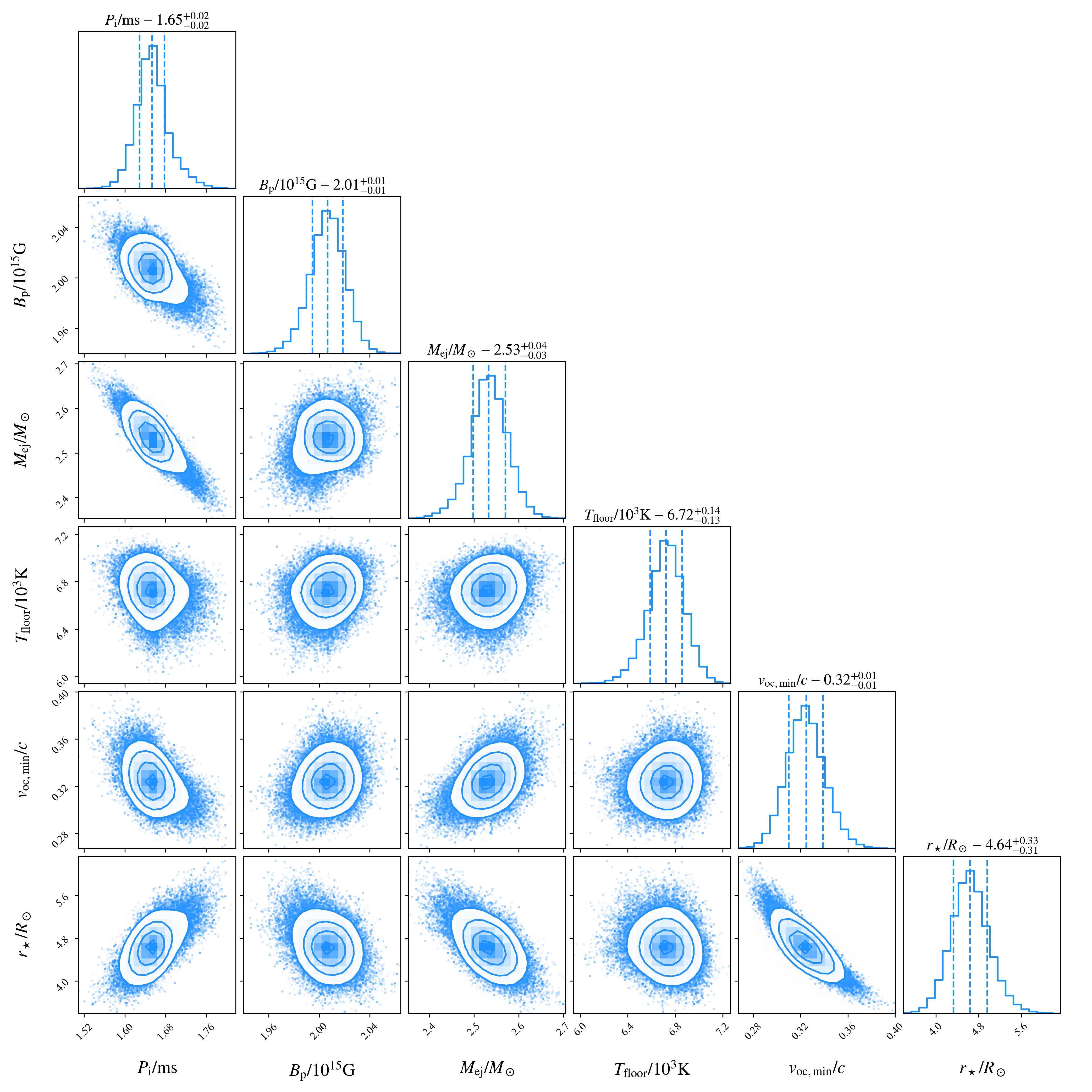}
\caption{{Posterior distributions of fits to SN\,2025kg. The medians and 1$\sigma$ ranges are labeled.}}
\label{fig:Corner}
\end{figure*}

{The priors and posteriors of our fitting parameters to SN\,2025kg are listed in Table \ref{tab:PriorPosterior}, while the corner plot of the posteriors is shown in Figure \ref{fig:Corner}.}

\begin{table}
    \centering
    \caption{{Prior and posterior of the fitting parameters. ${\rm U}(a,b)$ and ${\rm LogU}(a,b)$ represent uniform and log-uniform distributions between $a$ and $b$, respectively. The error bar includes $68\%$ credible intervals. }}
    \begin{tabular}{llc}
    \hline
    \hline
       Fitting Parameters  & Prior & Posterior  \\
       \hline
       $P_i/{\rm ms}$ & ${\rm U}(0.6,10)$ & $1.65^{+0.02}_{-0.02}$ \\
       $B_{\rm p}/{\rm G}$ & ${\rm LogU}(10^{13},10^{17})$ & $2.01^{+0.01}_{-0.01}\times10^{15}$ \\
       $M_{\rm ej}/M_\odot$ & ${\rm U}(0.5,20)$ & $2.53^{+0.04}_{-0.03}$ \\
       $T_{\rm floor}/{\rm K}$ & ${\rm LogU}(4000,10^4)$ & $6.72^{+0.14}_{-0.13}\times10^3$ \\
       $v_{\rm oc,min}/c$ & ${\rm U}(0.01,0.5)$ & $0.32^{+0.01}_{-0.01}$ \\
       $r_\star/R_\odot$ & ${\rm LogU}(10^{-1},10^4)$ & $4.64^{+0.33}_{-0.31}$ \\
       \hline
    \end{tabular}
    \label{tab:PriorPosterior}
\end{table}

\section{Bolometric Characteristics of SN\,2025kg} \label{app_sec:BolometricLightcurve}

\begin{figure}
\centering
\includegraphics[width=1\linewidth, trim = 55 23 105 60, clip]{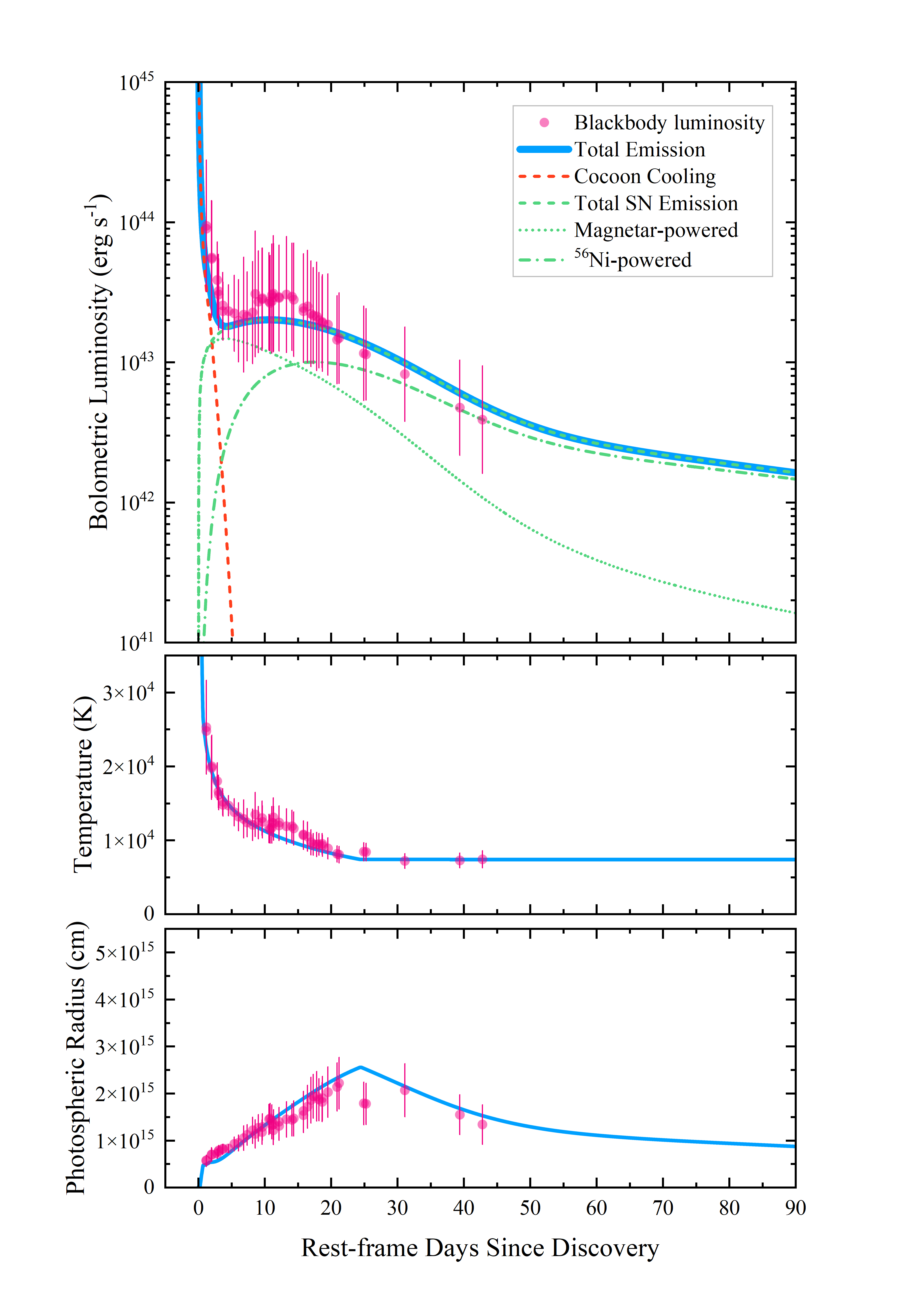}
\caption{Comparison of observed and simulated bolometric lightcurve (top panel), effective temperature (middle panel), and photospheric radius (bottom panel) for SN\,2025kg. The purple points and blue solid lines represent the observed and simulated combined bolometric characteristics of SN\,2025kg. In the top panel, the contributions from the outer cocoon cooling emission and SN emission are marked as orange and green dashed lines, respectively. In the SN emission, the inputs from the magnetar and $^{56}$Ni are represented by the green dotted and dashed-dotted lines.  }
\label{fig:BolometricEmission}
\end{figure} 

{Here, we show the observed and simulated bolometric characteristics for SN\,2025kg in Figure \ref{fig:BolometricEmission}. The observed bolometric lightcurve, effective temperature, and photospheric radius of SN\,2025kg are derived from blackbody fittings to the multi-band data using \texttt{SuperBol} \citep{Nicholl2018}. We directly simulate the bolometric lightcurve using the best fit to the multi-band lightcurves of SN\,2025kg, as presented in Section \ref{sec:BestFittings}. The simulated effective temperature and photospheric radius are determined by calculating the combined spectra of the outer cocoon cooling emission and SN emission at any given time, and then fitting these spectra to a blackbody curve.  }

\section{Afterglow Emissions} \label{app_sec:AfterglowModel}

{In this section, we briefly describe the afterglow model used to calculate the jet and inner cocoon afterglow lightcurves along the line of sight. Both the jet and inner cocoon are assumed to have top-hat structures.}

{The inner cocoon cooling emission mainly occurs within $\sim10\,{\rm days}$, during which the impact of the CSM on the dynamics of the cocoon can be negligible. In contrast, the afterglow emission on timescales of $\sim10^2-10^3{\rm days}$ requires careful consideration of the detailed dynamical evolution. The evolution of the Lorentz factor $\gamma_{\rm \{j,ic\}}$ of the jet or inner cocoon with the radius $r_{\rm \{j,ic\}}$ can be determined by the energy conservation law \citep{Nava2013,Zhang2018}, i.e.,}
\begin{equation}
\label{equ:Gamma}
\begin{split}
    &\frac{d\gamma_{\rm \{j,ic\}}}{dr_{\rm \{j,ic\}}}= \\
    &-\frac{(\gamma_{\rm eff,\{j,ic\}}+1)(\gamma_{\rm \{j,ic\}}-1)c^2\frac{dm_{\rm \{j,ic\}}}{dr_{\rm \{j,ic\}}}+\gamma_{\rm eff,\{j,ic\}}\frac{dE_{\rm ad,\{j,ic\}}}{dr_{\rm \{j,ic\}}}}{(m_{\rm \{j0,ic0\}}+m_{\rm \{j,ic\}})c^2+E_{\rm int,\{j,ic\}}\frac{d\gamma_{\rm eff,\{j,ic\}}}{d\gamma_{\rm \{j,ic\}}}},
\end{split}
\end{equation}
{where $m_{\rm \{j0,ic0\}}\approx E_{\rm \{j,ic\}}/\gamma_{\rm \{j0,ic0\}}c^2$ is the initial mass with $E_{\rm \{j,ic\}}$ and $\gamma_{\rm \{j0,ic0\}}$ being the kinetic energy and the initial Lorentz factor, $dm_{\rm \{j,ic\}}/dr_{\rm \{j,ic\}}$ is the evolution of the swept-up mass $m_{\rm \{j,ic\}}$, $dE_{\rm ad,\rm \{j,ic\}}/dr_{\rm \{j,ic\}}$ is the evolution of the adiabatic loss, $\gamma_{\rm eff,\{j,ic\}}=(\hat{\gamma}_{\rm \{j,ic\}}\gamma_{\rm \{j,ic\}}^2-\hat{\gamma}_{\rm \{j,ic\}}+1)/{\gamma_{\rm \{j,ic\}}}$,  $d\gamma_{\rm eff,\{j,ic\}}/dr_{\rm \{j,ic\}}=(\hat{\gamma}_{\rm \{j,ic\}}\gamma_{\rm \{j,ic\}}^2+\hat{\gamma}_{\rm \{j,ic\}}-1)/\gamma_{\rm \{j,ic\}}^2$, and $E_{\rm int,\{j,ic\}}$ is the internal energy, respectively. We take the post-shock adiabatic index $\hat{\gamma}_{\rm \{j,ic\}}$ from \cite{Peer2012}: $\hat{\gamma}_{\rm \{j,ic\}} = (5-1.21937z+0.18203z^2-0.96583z^3 + 2.32513z^4 - 2.39332z^5 + 1.07136z^6)/3$, where $z=\Theta/(0.24+\Theta)$ and}
\begin{equation}
    \Theta = \left(\frac{\gamma_{\rm \{j,ic\}}\beta_{\rm \{j,ic\}}}{3}\right)\left[\frac{\gamma_{\rm \{j,ic\}}\beta_{\rm \{j,ic\}}+1.07\gamma_{\rm \{j,ic\}}^2\beta_{\rm \{j,ic\}}^2}{1+\gamma_{\rm \{j,ic\}}\beta_{\rm \{j,ic\}}+1.07\gamma_{\rm \{j,ic\}}^2\beta_{\rm \{j,ic\}}^2}\right].
\end{equation}

{Considering an ambient density profile $\rho(r_{\rm \{j,ic\}})$, the evolution of the swept-up mass $m_{\rm \{j,ic\}}$ by the one-sided jet or cocoon can be calculated by}
\begin{equation}
    \frac{dm_{\rm \{j,ic\}}}{dr_{\rm \{j,ic\}}}=2\pi r_{\rm \{j,ic\}}^2(1-\cos\theta_{\rm \{j,ic\}})\rho(r_{\rm \{j,ic\}}),
\end{equation}
{where $\theta_{\rm \{j,ic\}}(r_{\rm \{j,ic\}})$ is the half-opening angle of the jet and cocoon. We adopt the lateral expansion of the jet and cocoon based on the jump conditions for oblique shocks of arbitrary 4-velocity \citep{Granot2012}, with the increased rate of}
\begin{equation}
    \frac{d\theta_{\rm \{j,ic\}}}{dr_{\rm \{j,ic\}}}\approx\frac{1}{\gamma_{\rm \{j,ic\}}^2\theta_{\rm \{j,ic\}}r_{\rm \{j,ic\}}},
\end{equation}
{The temporal evolution of the radius is determined by}
\begin{equation}
    \frac{dr_{\rm \{j,ic\}}}{dt}=\frac{\beta_{\rm \{j,ic\}} c}{1-\beta_{\rm \{j,ic\}}}.
\end{equation}

{One can express the evolution of the adiabatic loss in Equation (\ref{equ:Gamma}) as}
\begin{equation}
    \frac{dE_{\rm ad,\{j,ic\}}}{dr_{\rm \{j,ic\}}}=-(\hat{\gamma}_{\rm \{j,ic\}}-1)\left(\frac{3}{r_{\rm \{j,ic\}}}-\frac{1}{\gamma_{\rm \{j,ic\}}}\frac{d\gamma_{\rm \{j,ic\}}}{dr_{\rm \{j,ic\}}}\right)E_{\rm int,\{j,ic\}}.
\end{equation}
{The evolution of the internal energy can be described by}
\begin{equation}
\begin{split}
    \frac{dE_{\rm int,\{j,ic\}}}{dr_{\rm \{j,ic\}}}=&(1-\epsilon)(\gamma_{\rm \{j,ic\}}-1)c^2\cdot2\pi r_{\rm \{j,ic\}}^2(1-\cos\theta_{\rm \{j,ic\}})\rho-\\&(\hat{\gamma}_{\rm \{j,ic\}}-1)\left(\frac{3}{r_{\rm \{j,ic\}}}-\frac{1}{\gamma_{\rm \{j,ic\}}}\frac{d\gamma_{\rm \{j,ic\}}}{dr_{\rm \{j,ic\}}}\right)E_{\rm int,\{j,ic\}}.
\end{split}
\end{equation}
{The radiative efficiency can be expressed as $\epsilon=\epsilon_e\epsilon_{\rm rad}$ with $\epsilon_{\rm rad}=\min[1,(\gamma'_{e{\rm,m}}/\gamma'_{e{\rm,c}})^{p-2}]$, where $p$ is the power-law index of the electron Lorentz
factor distribution $\gamma'_e$, $\gamma'_{e{\rm,m}}=\epsilon_e(p - 2)m_p\gamma_{\rm \{j,ic\}}/[(p - 1)m_e]$, and $\gamma'_{e{\rm,c}}=6\pi m_ec/[\sigma_{\rm T}\gamma_{\rm \{j,ic\}} B'^2t'(1+Y)]$ are the minimum Lorentz factor and efficient cooling Lorentz factor of electrons, with the proton mass $m_p$, the electron mass $m_e$, the Thomson scattering cross section $\sigma_{\rm T}$, the magnetic
field behind the shock $B'= (32\pi\epsilon_B\rho)^{1/2}\gamma_{\rm \{j,ic\}} c$, and the Compton parameter $Y(\gamma'_e)$. We note the prime $'$ marks the comoving frame of the shock.}

{The synchrotron luminosity contributed by a differential element of a mass at the radius $R$ can be calculated analytically as}
\begin{equation}
    I'_{\nu'} \approx \frac{\int_{R_0}^R r_{\rm \{j,ic\}}^2[1-\cos\theta_{\rm \{j,ic\}}(r_{\rm \{j,ic\}})]\rho dr_{\rm \{j,ic\}}}{R^2[1-\cos\theta_{\rm \{j,ic\}}(R)]m_p}\frac{P'_{\rm \nu',max}S(\nu')}{1+Y(\nu')},
\end{equation}
{where $R_0$ is the initial radius, and the peak power is $P'_{\nu',{\rm max}} = m_ec^2\sigma_{\rm T}B'/12\pi q_e$ with $q_e$ being the electron charge. The dimensionless synchrotron spectrum $S(\nu')$ can be expressed as a broken-power-law function \citep{Sari1998,Sari1999,Gao2013,Zhang2018}, which is characterized by three broken frequencies $\nu_{\rm i}'=3q_eB'{\gamma'}_{e,{i}}^2/4\pi m_ec$, $i={\rm m},\,{\rm c},\,{\rm a}$, where $\gamma'_{e,{\rm a}}$ is the cooling Lorentz factor for synchrotron self-absorption, which is calculated using the blackbody method as described by \cite{Shen2009}. We include the synchrotron self-Compton process and model $Y(\gamma'_{e})$ following \cite{Jacovich2021} and \cite{McCarthy2024}.}

{Due to the aberration effect, photons emitted at $(R,\theta,\varphi)$ arrive at the observer with a time delay with respect to a photon emitted at $R=0$ by}
\begin{equation}
    t_{\rm obs} = (1+z)\int_0^R\frac{dR(1-\beta_{\rm \{j,ic\}}\cos\alpha)}{\beta_{\rm \{j,ic\}} c},
\end{equation}
{where $\cos\alpha=\cos\theta\cos\theta_{\rm v}+\sin\theta\sin\varphi\sin\theta_{\rm v}$ with $\theta_{\rm v}$ being the viewing angle relative to the jet/cocoon axis. By integrating the equal-arrival time surfaces, the flux density is given by}
\begin{equation}
    F_{\nu,{\rm \{j,ic\}}} =\frac{(1+z)R^2}{D_{\rm L}^2}\int\mathcal{D}^3I'_\nu(\nu_{\rm obs},t_{\rm obs},\theta,\varphi)\cos\alpha d\Omega
\end{equation}
{where the observed frequency is $\nu_{\rm obs}=\mathcal{D}\nu'/(1+z)$, the Doppler factor is $\mathcal{D} = 1/\gamma_{\rm \{j,ic\}}(1-\beta_{\rm \{j,ic\}}\cos\alpha)$, and $\Omega$ is the solid angle.}

{One can obtain multi-wavelength afterglow lightcurves of both the jet and inner cocoon separately by defining appropriate initial conditions and solving the above equations. The flux density from the counter-jet or counter-cocoon is calculated by setting $\theta_{\rm v}$ to $\theta_{\rm v}+\pi$. The total afterglow emission is obtained by combining the afterglow emissions of both the jet and inner cocoon.}

\bsp	
\label{lastpage}
\end{document}